# Deep Learning Accelerated Phase Prediction of Refractory Multi-Principal Element Alloys

Ali K. Shargh[1*], Christopher D. Stiles[1,2], Jaafar A. El-Awady[1*]

[1] Department of Mechanical Engineering, Johns Hopkins University, Baltimore, Maryland 21218, United States

[2] Research and Exploratory Development Department, Johns Hopkins University Applied Physics Laboratory, Laurel, Maryland 20723, United States

# Abstract

The tunability of the mechanical properties of refractory multi-principal-element alloys (RMPEAs) make them attractive for numerous high-temperature applications. It is well-established that the phase stability of RMPEAs control their mechanical properties. In this study, we develop a deep learning framework that is trained on a CALPHAD-derived database that is predictive of RMPEAs phases with high accuracy up to eight phases within the elemental space of Ti, Fe, Al, V, Ni, Nb, and Zr with an accuracy of approximately 90%. We further investigate the causes for the low out of domain performance of the deep learning models in predicting phases of RMPEA with new elemental sets and propose a strategy to mitigate this performance shortfall.

* Corresponding authors:
  Email addresses: ashargh1@jhu.edu (A. K. Shargh), jelawady@jhu.edu (J. A. El-Awady)

# Keywords

# 1. Introduction

Refractory multi-principal-element alloys (RMPEAs) are metallic alloys consisting of several principal elements with concentrations typically ranging from 5 to 35 atomic percentage (at.%) [1]. These alloys have recently garnered significant attention due to their exceptional properties exhibited by some RMPEAs, such as tensile strength, ductility, hardness, corrosion resistance, high-temperature oxidation resistance etc. [2–6]. This has sparked considerable interest in understanding how to design new RMPEAs with tailored properties.

It is widely accepted that the phase stability of RMPEAs can significantly impact their microstructure and, consequently, their mechanical properties. For instance, the face-centered cubic (FCC) phase has been found to enhance ductility, while the body-centered cubic (BCC) phase contributes to the strength of the RMPEAs [7]. Accordingly, dual-phase RMPEAs with FCC+BCC phases exhibit both high strength and ductility [8,9]. In addition, recent studies suggest that amorphous and intermetallic phases, such as Sigma, Laves, can also improve RMPEAs properties, including corrosion resistance, wear resistance, and elastic strain limit [10–12]. Therefore, accurately and efficiently predicting phase distributions is a crucial first step in designing novel RMPEAs with targeted properties. Given the vast compositional space of RMPEAs, experimental approaches alone are impractical for exploring the phase stability for the effective design of RMPEAs.

Several computational studies employed empirical parameters, first-principles density functional theory (DFT) calculations, and computer coupling of phase diagrams and thermochemistry (CALPHAD) to predict the phases of RMPEAs within specific regions of the design space [1,13–18]. While these approaches have demonstrated some success in predicting RMPEA phase stability, their utility of these approaches in exploring the entire compositional space of RMPEA is limited. This is due to high computational costs, low sensitivity to distinguish intermetallic phases, reliance on insufficient experimental data, and other factors.

To overcome the limitations of traditional computational methods, recent studies have increasingly utilized deep learning (DL) techniques to explore RMPEAs phase stability [19–28]. DL models excel at efficiently parameterizing and solving problems by learning complex patterns from training datasets. Consequently, several studies have used experimental datasets containing hundreds to over a thousand datapoints to train DL models and explore the design space [19–28]. For instance, Guo et al. [19] trained a convolutional neural network (CNN) model using only elemental composition as input features, with 914 experimental datapoints obtained by arc melting. They reported a prediction accuracy of 98% for solid solutions (SS) and amorphous (AM) phases, and 89% for intermetallic (IM) compounds. Similarly, Jain et al. [20] used an extra trees classifier with elemental compositions and seven thermo-physical parameters as input features, trained on 1120 experimental datapoints. They achieved a phase prediction accuracy of 89.3% for the classification of 6 different phase labels including FCC, FCC+IM, FCC+BCC, FCC+BCC+IM, BCC, and BCC+IM.

However, these studies have several limitations. First, the small experimental training datasets may not fully represent the design space, limiting the models' ability to capture the full complexity of RMPEA phase formation. Second, the number and resolution of predicted phases are often limited, with many studies using broad categories such as SS and IM as predictive labels, failing to predict more specific categories such as FCC, HCP, Laves, Sigma. Additionally, the models are typically restricted to classifying predefined phase combinations and cannot predict combinations beyond these predefined categories. Another issue is the uncertainties in synthesis routes, fabrication, and phase measurements can lead to inconsistent datapoints, which can incorrectly influence the DL models. Lastly, small datasets are often imbalanced, prompting the use of data augmentation techniques like under-sampling, over-sampling, and Synthetic Minority Over-sampling Technique (SMOTE). However, recent studies have highlighted challenges in integrating synthetic data from these augmentation techniques into experimental databases to accuracy improvement [29].

The challenge of limited experimental datasets can be partially addressed by combining computational predictions with DL models. One approach is to train DL models on databases created from CALPHAD calculations, using them as efficient surrogate models to quickly identifying RMPEA phases and accelerate the design space exploration [30–32]. For example, Vazquez et al. [32] trained a multi-layer perceptron (MLP) neural network with 32 features, each represented by three values calculated using weighted mean, weighted reduced mass, and weighted mean difference using 229,156 RMPEAs with 3-7 elements containing various combinations of Mn, Ni, Fe, Al, Cr, Nb, and Co, labeled with their expected phases as predicted from CALPHAD. They successfully predicted the fraction of various RMPEA phases including FCC, BCC, $C_{14}$ Laves, Sigma and IM with a coefficient of determination ($R^2$) greater than 0.96. Such $R^2$ indicates that the model accounts for over 96% of the variance in the data for predicting the fractions of different RMPEA phases. However, while R² is often used to assess model performance in regression problems, it does not fully capture the accuracy of phase predictions, particularly in terms of exactly matching all phase combinations present. Additionally, R² is sensitive to outliers in the data and does not provide information about the distribution or magnitude of residuals. All these factors highlight additional opportunities to improve predictions beyond what has been done in literature. Additionally, all trained DL models reported in literature report poor performance when predicting RMPEA phases outside of the training domain, such as with new elemental sets [32]. The root causes of this poor performance include the lack of comprehensive data and the models' inability to generalize beyond the specific compositions they were trained in. This highlights a limitation in the model's generalization capabilities and practical applicability to diverse materials.

Addressing these limitations will provide a substantial advance in the field of RMPEA design. Improving the accuracy and generalization capabilities of DL models requires innovative approaches to data generation, model training, and validation. Expanding the training datasets with more diverse and comprehensive data and refining the DL architectures could significantly enhance the predictive power and applicability of these models. This would enable more accurate and efficient exploration of the vast compositional space of RMPEAs, ultimately leading to the discovery and design of new alloys with tailored properties.

In this study, we first develop a framework that is trained on a large dataset of CALPHAD calculations and is capable of predicting a greater number of phases compared to earlier studies. Specifically, we will focus on a combination of 8 phase classifications including FCC, BCC, HCP, B2 (Ordered BCC), Laves (C14, C15 and C36), Sigma, Heusler, and Liquid. This data is then used to train a multi-label classification DL model to predict all possible combinations of these 8 phases, for compositions within the elemental space of Ti, Fe, Al, V, Ni, Nb, and Zr. The choice of the elements aligns with the mission of the Center on Artificial Intelligence for Materials in Extreme Environments (CAIMEE) to accelerate the design of RMPEAs for high-temperature applications. Next, we evaluate the DL model's performance in predicting phases for new elemental subsets that were not included in the initial trained data. We then propose an interpretable approach to enhance the effectiveness of DL models in predicting phases for elements outside of the training set.

The remaining sections of the paper are organized as follows. In the methods section, the details of the deep learning framework, including database generation, labeling, and the model architecture are discussed. The results section details the outcomes of our framework, including accuracy validation, the model's performance on new elemental subset, and strategies to address low performance. Finally, the conclusion section presents a summary of the study along with key findings.

# 2. Methods

## 2.1. Deep learning framework

### 2.1.1. Dataset preparation

The initial step in constructing our deep learning framework involves preparing a dataset that is labeled with both expected phases and the features that are crucial for predicting these phases. Given the expansive design space encompassing our seven elements of interest (Ti, Fe, Al, V, Ni, Nb, and Zr), we randomly sample it, generating 50,000 different compositions based on 3-5 elements. The fraction of each element can vary between 5-35%, with compositional increments of 1%. Generated samples are then labeled with 50 features chosen from the literature. Earlier studies [21,32–37] have confirmed that one or a combination of these parameters can be used to predict RMPEA phases. These 50 features include: mixing entropy $\Delta S_{mix}$, mixing enthalpy $\Delta H_{mix}$, $\Omega$, $\eta$, $k_1^{cr}$, $\Phi$, $\delta$, $\frac{E_2}{E_0}$, valence electron concentration VEC, $\Delta\chi$, $PFP_{FCC}$, $PFP_{BCC}$, $PFP_{HCP}$, $PFP_{B_2}$, $PFP_{Laves}$, $PFP_{Sigma}$, PSP, $\sigma_{\Delta H_{mix}}$, bulk modulus K, $\sigma_K$, melting temperature $T_m$, $\sigma_{T_m}$, $\sigma_{VEC}$, $\chi$, atomic number, atomic weight, period, group, families, Mendeleev number, L quantum number, miracle radius, covalent radius, Zunger radius, ionic radius, crystal radius, MB electronegativity, Gordy electronegativity, Mulliken electronegativity, Allred-Rockow electronegativity, first ionization potential, polarizability, boiling point, density, specific heat, heat of fusion, heat of vaporization, thermal conductivity, heat atomization and cohesive energy. The phase formation parameter (PFP) and phase separation parameter (PSP) are defined as the probability of forming different phases as well as phase separation respectively [38]. $\sigma$ is the standard deviation of different parameters. The parameter $\Phi$ is calculated using the publicly available code [36]. The parameters $\Delta S_{mix}$, $\Delta H_{mix}$, $\Omega$, $\eta$, $k_1^{cr}$, $\delta$, $\frac{E_2}{E_0}$, $\Delta\chi$ are defined in the Supplementary Table S1. The mean value of the remaining parameters from bulk modulus to cohesive energy are calculated from:

$$x_{avg} = \sum_{i=1}^{N} c_i x_i \qquad (1)$$

Wherein $c_i$ is the concentration of element i in atomic fraction, and $x_i$ is the parameter values for element i. The values of these parameters are publicly available [32]. RMPEAs in the initial dataset are then labeled with their expected phases using Python package Thermo-Calc [39] which is a widely used software that allows to predict various thermodynamic properties of materials such as equilibrium compositions and phase diagrams. We utilize the latest version of the thermodynamic database (i.e., TCHEA6) in our CALPHAD calculations. The temperature and pressure of the system are maintained at 850K and 1 bar, respectively. The choice of temperature aligns with the current application domain of the Center on Artificial Intelligence for Materials in Extreme Environments (CAIMEE). Also, 149 of our CALPHAD calculations were unsuccessful, leading to the removal of those compositions from the dataset. Such failures in CALPHAD calculations were reported in earlier studies as well [32].

### 2.1.2. Dataset engineering

To improve the training efficiency, we take the following steps before initiating the training process: 1) We normalize all feature values to the range (0,1) for the entire dataset, using the minimum and maximum values of each feature. Then, we examine histograms (bin size = 0.1) for all features and remove datapoints in bins with low intensity (intensity < 20). This results in the removal of 508 compositions from the dataset. This step is effective in eliminating datapoints with features that deviate significantly from the remaining of the dataset and helps to have feature distributions that are more uniform.

2) We calculate the Pearson correlation coefficient (PCC) between each pair of the 50 initial features and remove those that are strongly correlated ($|PCC| > 0.9$), retaining only one feature from each correlated pair. Figure 1 illustrates the PCC matrix, highlighting highly correlated feature pairs. From this matrix, we eliminate 16 features including: $\frac{E_2}{E_0}$, $\Delta\chi$, atomic weight, period, Mendeleev number, ionic radius, crystal radius, Mulliken electronegativity, Allred-Rockow electronegativity, first ionization potential, polarizability, heat of fusion, heat of vaporization, thermal conductivity, heat atomization and cohesive energy.

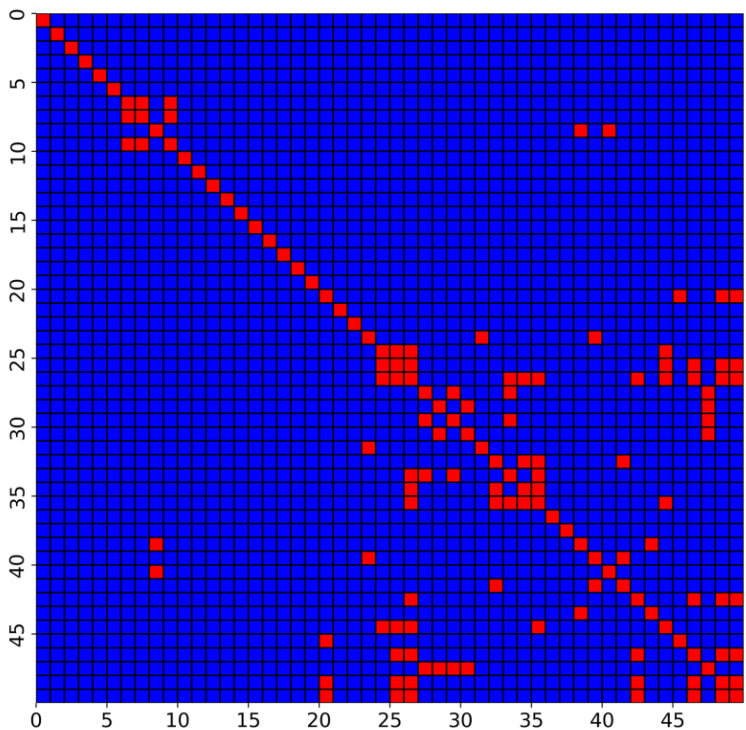

**Figure 1: Feature correlation matrix of the dataset. Pearson correlation coefficient matrix for the 50 input features (IDs 0 to 49). Note that those pairs that are highly correlated (PCC>0.9) are colored with red. From each highly correlated pair features, one is removed for training efficiency.**

### 2.1.3. Deep neural network

To achieve the goal of predicting RMPEA phases in this study, we opt for the multi-layer perceptron (MLP) neural network. MLP is recognized for its proficiency in predicting output properties from input features by learning the salient characteristics of the training database during the training process. The architecture of the MLP network of our deep learning framework is shown in Figure 2. In this architecture, the hyperbolic tangent (i.e., tanh) is used as the activation function in all layers, except the final one. This choice aids the network in learning complex nonlinear correlations. Given that the phase prediction problem is framed as a multi-label classification in this study, the sigmoid activation function is employed for the final layer. This choice enables each output neuron to independently predict the presence or absence of each of the 8 phases of interest. We use binary cross-entropy loss function with Adam optimizer [40] to train the MLP network. A 10-fold cross-validation is carried out to prevent overfitting. Initially, 10% of the labeled dataset is randomly chosen as the testing dataset to assess the final well-trained model. The remaining labeled data is divided into 10 subsets, and the model's performance is validated on one subset while trained on the remaining 9 subsets. The accuracy of the model is then averaged over ten folds, providing a robust evaluation of its generalization capabilities.

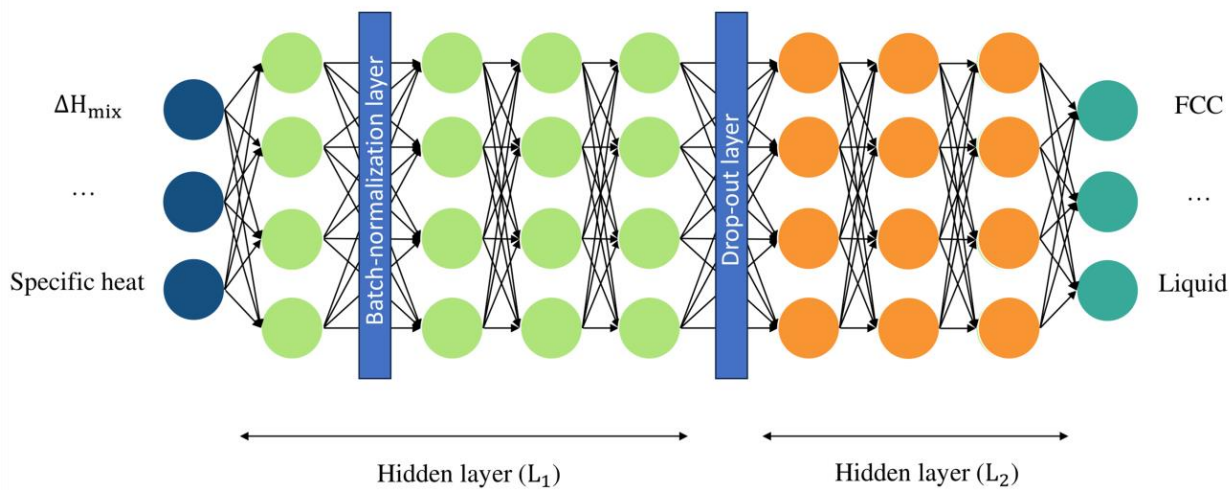

**Figure 2: The architecture of the MLP network in this study. In this architecture, the number of the neurons are fixed for all hidden layers, one layer of drop-out is added in between hidden layer groups shown with $L_1$ and $L_2$, and one layer of batch normalization is added after the hidden layer group shown with $L_1$.**

It is notable that Bayesian optimization technique [41,42] is used to fine-tune hyper parameters of the network including: the number of hidden layers (i.e., $L_1$ and $L_2$ shown in Figure 2), the number of neurons in each layer, batch size, learning rate, the presence or absence of BN and drop-out layers, and the drop-out rate value. Bayesian optimization, known for its effectiveness in optimizing complex functions, is particularly well-suited for determining the optimal set of hyperparameters based on the performance of MLP network in predicting the phases of RMPEAs on the validation dataset. The fine-tuned architecture of different MLP network used in this study is summarized in Table 1. The optimization step initiates with a random space comprising 40 points and continues through 80 iterations. It should be noted that the MLP network architecture is implemented based on PyTorch.

Table 1: The fine-tuned architecture of different MLP network used in this study. Note that a value of 0 or 1 for Drop-out and batch normalization (BN) layers indicates their absence or presence respectively.

| Training database | Drop_out | Drop_out rate | BN | Batch size | $L_1$ | $L_2$ | Neurons | Learning_rate |
|---|---|---|---|---|---|---|---|---|
| All with 34 features | 0 | - | 0 | 309 | 6 | 4 | 85 | 0.00039 |
| Ti-free with 34 features | 1 | 0.539 | 0 | 345 | 10 | 0 | 44 | 0.00043 |
| All with 27 features | 0 | - | 0 | 313 | 6 | 7 | 36 | 0.00172 |
| Ti-free with 27 features | 1 | 0.539 | 0 | 345 | 10 | 0 | 44 | 0.00043 |

# 3. Results

## 3.1. Validation of framework accuracy

Figure 3(a) shows the learning curve of the MLP network (architecture detailed in Figure 2 and Table 1 of section 2.). The network was trained on the RMPEA dataset to predict eight phase classifications: FCC, BCC, HCP, B2 (Ordered BCC), Laves (C14, C15, C36), Sigma, Heusler, and liquid. The model's performance is evaluated using ten-fold cross-validation, with the solid lines representing the average training (black) and validation (red) accuracies across the folds. The shaded areas surrounding the lines depict the accuracy variation among individual folds. In this study, accuracy is defined as the proportion of RMPEAs for which all eight associated phases are correctly predicted.

During the first 50 epochs, the training accuracy rapidly improves before plateauing, while the validation accuracy follows a similar trend, reaching its maximum value at epoch 211, as shown in Figure 3(a). This suggests that the DL model has effectively captured the correlations between the input features and the RMPEA phases. The best-performing model at epoch 211 achieves accuracies of 0.9490 (±0.009), 0.8951 (±0.003), and 0.8992 (±0.004) on the training, validation, and testing datasets, respectively. The values in parentheses represent the standard deviation across the ten folds. Although training was continued for 3,000 epochs, no further improvement in the validation accuracy was observed beyond epoch 211.

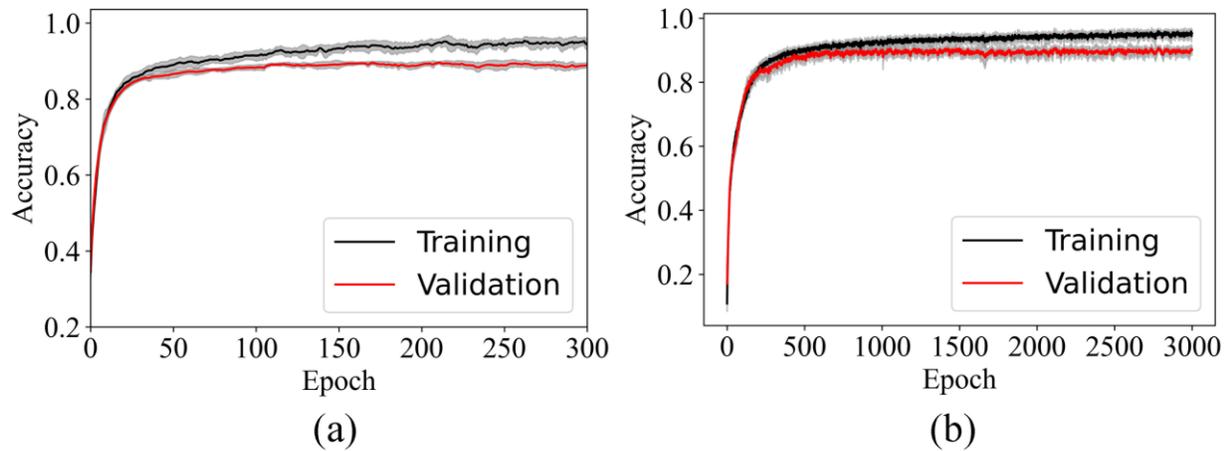

**Figure 3:** Learning curve of the MLP model for predicting the combination of eight possible phases for RMPEAs with the phase classifications include FCC, BCC, HCP, B2 (Ordered BCC), Laves (C14, C15 and C36), Sigma, Heusler and Liquid. a) The model trained on the complete dataset (Ti, Fe, Al, V, Ni, Nb, and Zr). The curves are truncated after 300 epoch with the best performance observed at epoch 211. b) The model trained on a Ti-free dataset. The curves are truncated at epoch 3000 with the best performance observed at epoch 2784. In both figures the solid lines represent the average training (black) and validation (red) accuracies across 10 folds, while the shaded gray areas depict the accuracy variation among individual folds.

While the overall accuracy of the DL model is 0.9 in terms of predicting all eight phases simultaneously, it is interesting to further assess the performance of the DL model in predicting individual phases. For this we define three parameters namely, precision, recall, and the F1 score as follows:

$$\text{Precision} = \frac{\text{TP}}{\text{TP+FP}} \tag{2}$$

$$\text{Recall} = \frac{\text{TP}}{\text{TP+FN}} \tag{3}$$

$$\text{F1} = \frac{2 \times (\text{Precission} \times \text{Recall})}{\text{Precission+Recall}} \tag{4}$$

where TP is the number of the true positive predictions (correctly predicted presence of a phase), FN is the number of false negative predictions (incorrectly predicted absence of a phase), and FP is the number of false positive predictions (incorrectly predicted presence of a phase). The summary of the performance of the DL model for predicting each phase of the testing dataset separately is reported in Table 2. The 'Ratio' column in the table indicates the proportion of the training dataset containing the phase shown in each row.

**Table 2: Performance metrics of the DL model in predicting the presence of individual phases in the testing dataset (4,933 alloys, fold=1). The model was trained on the complete dataset to simultaneously predict the presence of all eight phases.**

|  | Ratio | False Positive (FP) | False Negative (FN) | True Positive (TP) | True Negative (TN) | Precision (Eq. (2)) | Recall (Eq. (3)) | F1 (Eq. (4)) |
|---|---|---|---|---|---|---|---|---|
| **FCC** | 0.01 | 7 | 3 | 46 | 4877 | 0.87 | 0.94 | 0.9 |
| **BCC** | 0.59 | 34 | 61 | 2875 | 1963 | 0.99 | 0.98 | 0.98 |
| **HCP** | 0.05 | 40 | 17 | 256 | 4620 | 0.86 | 0.94 | 0.9 |
| **B2** | 0.74 | 89 | 57 | 3576 | 1211 | 0.98 | 0.98 | 0.98 |
| **Laves** | 0.82 | 25 | 37 | 4107 | 764 | 0.99 | 0.99 | 0.99 |
| **Sigma** | 0.19 | 27 | 38 | 898 | 3970 | 0.97 | 0.96 | 0.97 |
| **Heusler** | 0.29 | 19 | 11 | 1463 | 3440 | 0.99 | 0.99 | 0.99 |
| **Liquid** | 0.05 | 21 | 18 | 280 | 4614 | 0.93 | 0.94 | 0.93 |

Table 2 demonstrates that the F1 score for all phases exceeds 0.9, highlighting the model's high level of accuracy and balance between precision and recall in predicting individual RMPEA phases. The model excels in predicting phases such as BCC, B2, Laves, and Heusler, achieving F1 scores above 0.98. However, the lower F1 score for phases such as: FCC, HCP and Liquid may be attributed to their underrepresentation in the training dataset, as indicated by the 'Ratio' column in

Table 2. Underrepresentation of certain classes in the training data is a common challenge in data science, as it can lead to biased models that struggle to learn the distinguishing features of minority classes effectively [43–45]. Consequently, the model would find it more challenging to accurately predict these phases compared to others that are more abundant in the training dataset.

Despite the class imbalance, the DL model demonstrates strong overall performance, with all phases achieving F1 scores above 0.9. This suggests that the model has effectively learned to identify the key features associated with each phase, enabling accurate predictions even for less frequently encountered phases. The high F1 scores across all phases underscore the model's robustness and its ability to generalize well to the testing dataset, which is essential for its practical application in predicting RMPEA phases.

## 3.2. Out of domain performance of the trained DL model

Investigating the out-of-domain performance (generalization) of the DL model in predicting RMPEAs phases with new elemental components not in the training set is of general interest, as it demonstrates the model's ability to make accurate predictions in novel compositional regions. This generalizability would enhance the practical application of the DL model, highlighting its adaptability in handling elements not contained in the training set, enabling the exploration and discovery of new RMPEAs. To evaluate the DL model performance in out-of-domain predictions, we train and evaluate a new model on a modified database where all Ti containing alloys have been removed. The model performance is then tested on the Ti-containing alloys. The learning curve of the training process for 3000 epochs is shown in Figure 3(b). Similar to the earlier model, the accuracy of both training and validation datasets improves until reaching a plateau. Specifically, the accuracy of the validation dataset containing Ti-free alloys reaches its maximum value of 0.9074($\pm$0.005) at epoch 2784, and the corresponding value for the training dataset is 0.9525(±0.009). However, the accuracy of the model on the testing dataset containing Ti-alloys decreases to 0.2235(±0.008). This large decline in accuracy on the testing dataset indicates that the current model performs poorly when predicting phases of Ti-alloys.

It is important to note that while the accuracy on the Ti-containing test set is low, it is still significantly higher than the probability of randomly predicting all 8 labels correctly. With each label having a 0.5 chance of being correctly predicted by random change, the probability of accurately predicting all 8 labels simultaneously is $(0.5)^8 = 0.004$.

Table 3 summarizes the performance of the DL model, trained only on the Ti-free alloys dataset, in predicting the individual phases of Ti-alloys. It is observed that the model has a high F1 score for predicting BCC and Laves phases, possibly due to a higher number of RMPEAs with those phases in the training dataset. Interestingly, despite the B2 phase also being well-represented in the training dataset, its corresponding F1 score is not as high. Moreover, the Heusler phase prediction achieves a high F1 score despite having a low representation in the training dataset. Those observations suggest that the representation of a phase in the training set (i.e., the number of RMPEAs containing that phase) is not the sole factor affecting the model's predictive accuracy, and other factors may play important roles. In section 3.3., we delve deeper into this aspect to better understand the factors influencing the model's performance and suggest a method for improvement.

**Table 3: Performance metrics of the DL model in predicting the presence of individual phases in the Ti-containing testing dataset (34,090 alloys, fold=1). The model was trained on the Ti-free dataset to simultaneously predict the presence of all eight phases and tested on Ti-alloys.**

|  | Ratio | FP | FN | TP | TN | Precision | Recall | F1 |
|---|---|---|---|---|---|---|---|---|
| **FCC** | 0.006 | 41 | 450 | 84 | 33515 | 0.67 | 0.16 | 0.25 |
| **BCC** | 0.62 | 3288 | 2771 | 17017 | 11014 | 0.84 | 0.86 | 0.85 |
| **HCP** | 0.05 | 4241 | 1453 | 234 | 28162 | 0.05 | 0.14 | 0.08 |
| **B2** | 0.58 | 2639 | 13064 | 14794 | 3593 | 0.85 | 0.53 | 0.65 |
| **Laves** | 0.90 | 5031 | 714 | 26435 | 1910 | 0.84 | 0.97 | 0.9 |
| **Sigma** | 0.20 | 2013 | 2963 | 3534 | 25580 | 0.64 | 0.54 | 0.59 |

| Heusler | 0.37 | 4587 | 1208 | 7796 | 20499 | 0.63 | 0.87 | 0.73 |
| Liquid | 0.08 | 2991 | 293 | 889 | 29917 | 0.23 | 0.75 | 0.35 |

## 3.3. Enhancement of out of domain Ti-alloy prediction

As it was discussed in section 3.2., the DL model faces challenges in predicting the phases of Ti-alloys when trained only on Ti-free alloys. To enhance the model performance, our approach involves identifying the most common phase combination in the testing dataset and subsequently improving the accuracy of the datapoints with that specific phase combination.

Table 4: Top 10 phase combinations ranked in order for both the training (fold=1) and the testing datasets. The accuracy in predicting the different phase combinations is also reported. The model was trained on the Ti-free dataset to simultaneously predict the presence of all eight phases and tested on Ti-alloys

| Training dataset | | | Testing dataset | | |
| --- | --- | --- | --- | --- | --- |
| Phase Combinations | Ratio | Accuracy | Phase Combinations | Ratio | Accuracy |
| BCC+B2+Laves | 0.14 | 0.97 | BCC +B2+Laves | 0.24 | 0.26 |
| BCC+Laves | 0.13 | 0.98 | B2+Laves | 0.14 | 0.30 |
| B2+Laves | 0.10 | 0.98 | BCC +B2 | 0.08 | 0.054 |
| B2+Laves+ Heusler | 0.08 | 0.98 | BCC +Laves | 0.07 | 0.29 |
| BCC+B2+Laves+Heusler | 0.07 | 0.92 | B2+Laves+ Heusler | 0.06 | 0.59 |
| BCC +Laves+Heusler | 0.07 | 0.98 | B2+Laves+Sigma | 0.05 | 0.14 |
| B2+Laves+Sigma | 0.05 | 0.99 | BCC +B2+Heusler | 0.03 | 0.18 |
| BCC +HCP+Laves | 0.03 | 0.96 | BCC+B2+Laves+Heusler | 0.03 | 0.19 |
| BCC +Laves+Liquid | 0.03 | 0.97 | HCP+B2+Laves | 0.02 | 0 |
| Laves+Sigma | 0.02 | 0.99 | BCC+B2+Laves+Sigma | 0.01 | 0 |

Table 4 shows the top 10 phase combinations for the testing dataset. Almost a quarter of the testing datasets (24%) have a combination of BCC+B2+Laves phases, with the model accuracy of predicting this combination being 0.26. Interestingly, a large portion of the training dataset also has the combination of BCC+B2+Laves phases. This suggests that the low performance of the DL model in predicting Ti-alloys with BCC+B2+Laves phases is not solely due to training on a limited number of RMPEAs with BCC+B2+Laves phases. We hypothesize that the low performance could stem from one or both of the following factors:

1. The features associated with the Ti-alloys having BCC+B2+Laves phases might not overlap with the feature distribution of Ti-free alloys, making them outliers in the feature space.
2. The feature patterns mapped to BCC+B2+Laves phases that are learned by the DL model from the Ti-free alloys during the training process are different from those of Ti-alloys with BCC+B2+Laves phases. In other words, the model's learned representation of the BCC+B2+Laves phase combination, based on the features of Ti-free alloys, does not accurately capture the features of Ti-alloys with the same phase combination.

To investigate these hypotheses and improve the model's performance on out-of-domain Ti-alloy phase prediction, we propose an approach that involves analyzing the feature distributions of Ti-free and Ti-containing alloys, as well as examining the model's learned feature-phase mappings. By understanding the differences in feature patterns between the two datasets and how the model's learned representations differ for Ti-free and Ti-containing alloys, we aim to develop strategies for enhancing the model's ability to accurately predict the phases of out-of-domain alloys. The following sections will delve into the details of our proposed approach and present the results of our analysis.

### 3.3.1. Comparative analysis of feature distributions: Ti-Free vs. Ti-Alloys

To investigate our first hypothesis, we compare the feature distribution of Ti-free alloys having a combination of BCC+B2+Laves phases and both correctly and falsely classified Ti-alloys with BCC+B2+Laves phases. Figure 4 shows the distribution for two representative features, while the remaining 32 features are shown in Supplementary Figure S1-S7. The red arrows in the figures highlight specific ranges considered as 'outliers' in the feature space. Upon comparing the feature distribution of Ti-free alloys and falsely classified Ti-alloys, it is evident that for some features, the outlier ranges are absent from the feature distribution of Ti-free alloys. This suggests that the features of some falsely classified Ti-alloys might not overlap with the feature distribution of Ti-free alloys.

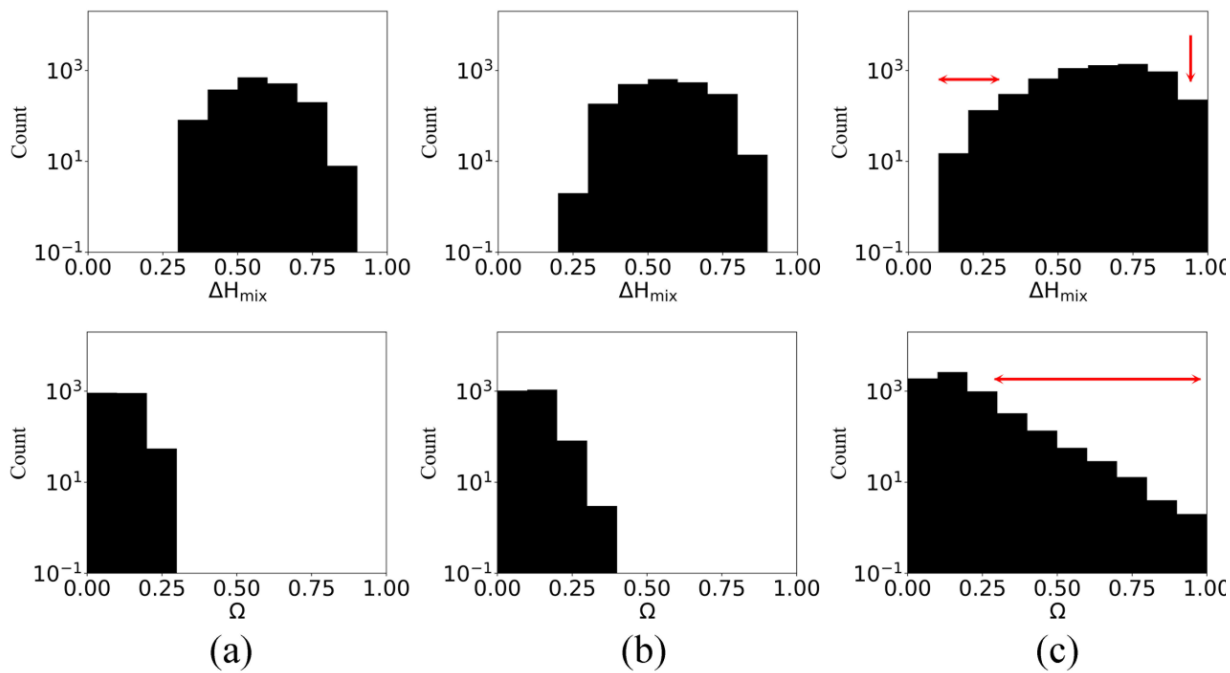

**Figure 4: Comparison of the feature distributions (range: 0-1) for alloys with BCC+B2+Laves phases. a) Normalized feature distributions for Ti-free. b) Normalized feature distributions for Ti-containing alloys correctly classified. c) Normalized feature distributions for Ti-containing alloys falsely classified. Only two representative features of $\Delta H_{mix}$, and $\Omega$ are shown; see Supplementary Figure S1-S7 for the remaining 32 features. Red arrows indicate outlier ranges.**

To further quantify the prevalence of outliers, the distribution of the number of outlier features for falsely classified Ti-alloys is shown in Figure 5. Nearly 70% of the falsely classified Ti-alloys exhibit only 0 to 2 outlier features, implying that the majority of the 6,106 falsely classified Ti-alloys have feature values that largely overlap with the feature distribution of Ti-free alloys.

Consequently, the first hypothesis is ruled out as the primary explanation for the model's low performance on Ti-alloys. The fact that a substantial majority of falsely classified Ti-alloys have feature values that overlap with the Ti-free alloy distribution suggests that other factors, such as the model's learned feature-phase mappings, may play a more significant role in the poor out-of-domain performance. In section 3.3.2., we will investigate the second hypothesis, which focuses on the differences in the model's learned representations for Ti-free and Ti-containing alloys.

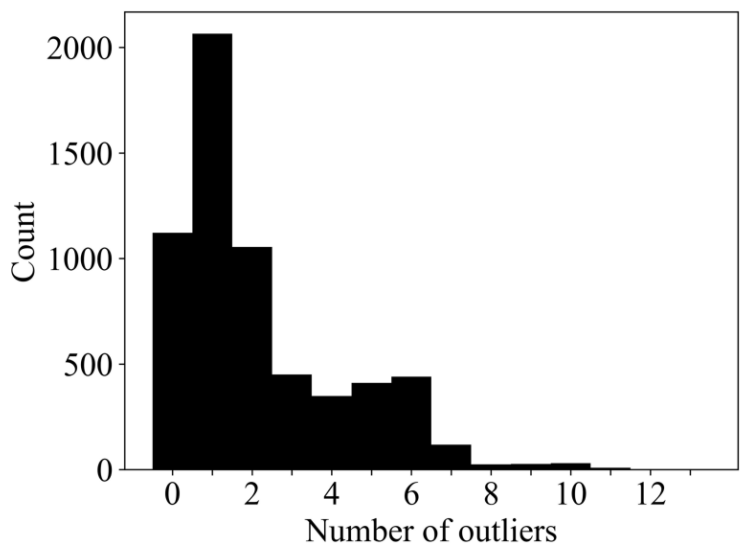

**Figure 5: Outlier feature counts for misclassified Ti-alloys with BCC+B2+Laves phases. Distribution of outlier feature counts for falsely classified Ti-alloys with BCC+B2+Laves phases.**

### 3.3.2. Distinctive feature patters: BCC+B2+Laves phases in Ti-Free vs. Ti-Alloys

To investigate our second hypothesis, we use the 'NearestNeighbor' module from the scikit-learn Python library [46] to identify the closest Ti-free alloy (in terms of feature similarity) for each of the 6,106 falsely classified Ti-alloys with BCC+B2+Laves phases. The nearest neighbor approach was selected to quantify feature similarity and provide insights into potential overlaps or distinctions between the falsely classified Ti-alloys and their closest Ti-free counterparts. We compare the phase combination of each Ti-alloy with its closest Ti-free alloy and record the number of incorrectly predicted phases among the eight phase labels for each sample. Interestingly, only 1543 of the 6106 closest Ti-free alloys are unique, indicating that many Ti alloys have similar closest Ti-free counterparts. Additionally, we calculated the average distance between each Ti-alloy and its closest Ti-free alloy based on the value derived from the 'NearestNeighbor' module, finding it to be 0.4307.

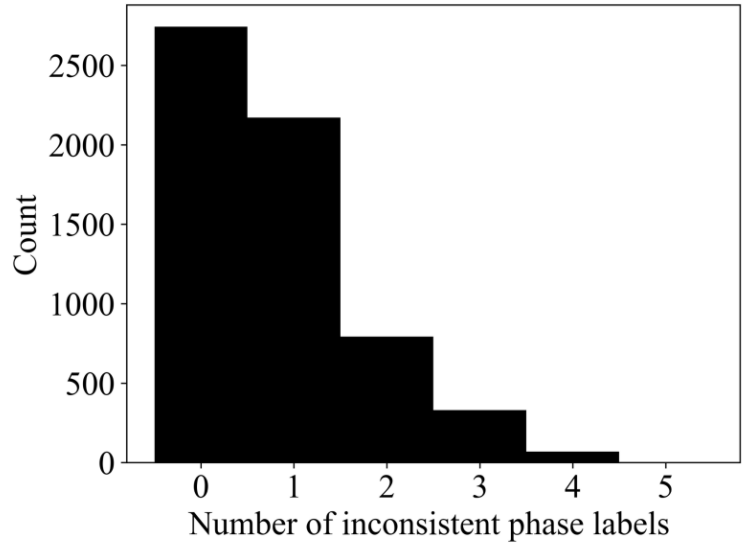

**Figure 6: Inconsistent phase counts in misclassified Ti-alloys and closest Ti-Free alloys. Distribution of the number of incorrectly predicted phases of falsely classified Ti-alloys with BCC+B2+Laves phases and the true phase labels of their closest Ti-free alloys from the training dataset**

The distribution of the number of incorrectly predicted phases of Ti-alloys with BCC+B2+Laves phases and true phase labels of their closest Ti-free alloys is shown in Figure 6. Remarkably, over 80% of the falsely classified Ti-alloys have only 0 or 1 incorrectly predicted phases when compared to the true phase labels of their closest Ti-free alloys. This finding confirms that the current DL model has learned to map feature combinations associated with falsely classified Ti-alloys into different phases than BCC+B2+Laves. Consequently, this result suggests a potential avenue for improving the performance of DL model on Ti-alloys: augmenting the Ti-free training dataset by including Ti-free alloys with BCC+B2+Laves phases that exhibit feature combinations similar to those of the 6,106 falsely classified Ti-alloys. In section 3.3.3., we explore the feasibility of augmenting the Ti-free training dataset and evaluate its impact on the model's performance.

**3.3.3. Augmentation of the Ti-free database to improve the predicting accuracy of Ti-alloys**

As discussed in section 3.3.2., we hypothesize that augmenting the training dataset by incorporating Ti-free alloys with BCC+B2+Laves phases and features similar to the 6,106 falsely classified Ti-alloys will enhance the performance of the DL model. To test this hypothesis, we follow these steps:

1) Using the DL model trained on all 7 elements in section 3.1., we explore the design space of Ti-free alloys, encompassing 3492741 unique compositions with an increment of 0.01 to identify Ti-free candidates with BCC+B2+Laves phases.

2) From this refined design space, we identify the nearest neighbour (based on feature similarity) of each of the 6,106 falsely classified Ti-alloys with BCC+B2+Laves phases.

3) The phases of the newly identified closest Ti-free alloys are validated using CALPHAD. Those that are not classified as BCC+B2+Laves will be removed.

4) From the original training dataset, we remove the closest Ti-free alloys to the 6,106 falsely classified Ti-alloys that are not BCC+B2+Laves. This step ensures that the closest Ti-free alloys to the falsely classified Ti-alloys with BCC+B2+Laves phases in the modified training dataset all exhibit BCC+B2+Laves phases.

5) Finally, the new samples obtained in step 3 are added to the training dataset, and the training process is repeated.

It should be noted that in the following analysis, the input feature lists were reduced to 27 features by ignoring the seven phase formation parameters (PFP) and phase separation parameter (PSP), defined in section 2. This was conducted to decrease the computational time needed to label the entire design space while not significantly affecting the testing accuracy of the DL models trained on all elements or Ti-free alloys, as shown in Table 5.

**Table 5: Accuracy comparison of DL models trained on different datasets and feature sets. Reducing input features from 34 to 27 minimally impacts accuracy. Augmenting the Ti-free dataset with strategically selected Ti-free alloys improves testing accuracy by 7%, highlighting the effectiveness of targeted dataset modifications for enhancing out-of-domain predictions.**

| Model description | Training accuracy | Validation accuracy | Testing accuracy |
| --- | --- | --- | --- |
|  |  |  |  |
| Trained on all elements with 34 features | 0.9490(±0.009) | 0.8951(±0.003) | 0.8992(±0.004) |
| Trained on Ti-free alloys with 34 features | 0.9525(±0.009) | 0.9074(±0.005) | 0.2235(±0.008) |
| Trained on all elements with 27 features | 0.9339(±0.011) | 0.8727(±0.006) | 0.8736(±0.004) |
| Trained on Ti-free alloys with 27 features | 0.9354(±0.009) | 0.9069(±0.011) | 0.2325(±0.011) |
| Trained on Ti-free alloys (modified) | 0.9452(±0.012) | 0.8652(±0.013) | 0.3018(±0.011) |

Following the augmentation steps, we remove 1384 Ti-free alloys that were not labeled as BCC+B2+Laves phases from the original training dataset and add 1586 new Ti-free alloys with BCC+B2+Laves phases. This modification increases the ratio of the number of closest Ti-free alloys to those falsely classified Ti-alloys with BCC+B2+Laves phases from 0.16 to 0.82. Upon completing the training process, the accuracy of the testing dataset is calculated as 0.3018 (±0.011), as reported in Table 5, representing an improvement of 7% increase over the original accuracy of 0.2325 (±0.011). This improvement in accuracy suggests that the proposed method for augmenting the accuracy of the DL model trained on Ti-free alloys is effective.

However, we acknowledge the importance of the average distance between the falsely classified Ti-alloys with BCC+B2+Laves phases and their closest Ti-free alloys as another key parameter. Our calculations show that the average distance decreases from 0.4307 for the original dataset to 0.3827 for the modified dataset. While this represents a clear improvement, future attempts should aim to further decrease this parameter to achieve more significant improvements in the model's performance. Analyzing the distribution of distances between falsely classified Ti-alloys and their closest Ti-free alloys could provide valuable insights into the impact of this parameter on the model's accuracy.

# 4. Conclusion

RMPEAs represent a unique class of metallic alloys with remarkable tunable mechanical properties. It has been demonstrated that the phase stability of RMPEAs plays a key role in controlling their mechanical properties. While phase prediction is crucial for designing novel RMPEAs with desired properties, current available methods face challenges to fully address it owing to the computational costs, insufficient experimental data, along with the low number and resolution of the predicted phases. In this study, we presented a deep learning framework to predict the RMPEA phases with high accuracy. Below are the conclusive points from our study:

- Considering the large compositional space of RMPEAs, it is not feasible to explore the whole design space using experiments or even lower fidelity computational approaches including CALPHAD. Here, we randomly sampled 50,000 points within the design space defined by our targeted elements: Ti, Fe, Al, V, Ni, Nb and Zr using CALPHAD. Using the prepared database, we successfully trained a DL model that can predict the phases of RMPEAs from 34 input physical descriptors, achieving an accuracy of approximately 90%. We selected those 34 descriptors based on their success in predicting the phases of RMPEAs from prior experiments and computational studies.
- One concern raised in the earlier studies was the low out of domain performance of the DL models in prediction of RMPEA phases with new elemental sets. We evaluated the performance of our DL model on Ti-alloys while trained on Ti-free alloys to address the issue. The accuracy of the trained network on Ti-alloys was low as anticipated, approximately 22%, while the features of the majority of falsely classified Ti-alloys overlap with the feature distribution of Ti-free alloys. Interestingly, our results unveiled that the feature patterns mapped to BCC+B2+Laves phases that are learned by the DL model from the Ti-free alloys during the training process are different from those of Ti-alloys with BCC+B2+Laves phases and thus the feature-phase mapping is not entirely unique for different alloys.

- An avenue to enhance the DL model's performance on Ti-alloys, when trained on Ti-free alloys, involves augmenting the Ti-free training dataset. This includes the addition of Ti-free alloys with phase labels and features that are closely resembling those of falsely classified Ti-alloys. Using this approach, we observed a 7% improvement from the original 22% performance of our DL model. However, it is recognized that the average distance between falsely classified Ti-alloys and their closest Ti-free alloys from the training dataset is another significant parameter that constrains the extent of accuracy improvement. As a result, further efforts are required to continue refining the out of domain performance by focusing on decreasing this parameter.
- The framework we developed in this study can readily be retrained and generalized to predict the phases of RMPEAs with other elements and phases. It is worth noting that the accuracy of CALPHAD is not perfectly predictive of physical experiments. Therefore, a future strategy involves combining this approach with experimental data to enhance the model's decision-making accuracy. In our upcoming study, we continue to pursue this approach. Simultaneously, recognizing the limitations of available experimental databases, CAIMEE aims to provide public experimental database for the community as one of its missions.

## Author contributions

A.K.S., C.D.S., and J.E.A contributed to the conception of the study. A.K.S., C.D.S., and J.E.A contributed to the setup and design of the framework. A.K.S. developed automatic dataset generation pipeline, integrated the framework and trained the model. A.K.S., C.D.S., and J.E.A contributed to the analyzing of the results and providing conclusive points. A.K.S., C.D.S., and J.E.A contributed to the writing and revising of the manuscript.

# Acknowledgment

This work was supported by the Army Research Laboratory and was accomplished under Cooperative Agreement Number W911NF-22-2-0014. The authors gratefully acknowledge internal financial support from the Johns Hopkins University Applied Physics Laboratory's Independent Research & Development (IR&D) Program. The authors acknowledge S. Joseph Poon and Jie Qi of the University of Virginia for their fruitful discussion on the calculation of phase formation and phase separation parameters. Computational resources were provided by the Advanced Research Computing at Hopkins (ARCH).

# Declaration of competing interest

All authors declare no financial or non-financial competing interests.

# Data availability

All data generated, used and/or analyzed during the current study are available from the corresponding author on request.

# Supplementary material for

# Deep Learning Accelerated Phase Prediction of Refractory Multi-Principal Element Alloys

Ali K. Shargh[1*], Christopher D. Stiles[1,2], Jaafar A. El-Awady[1*]

[1] Department of Mechanical Engineering, Johns Hopkins University, Baltimore, Maryland 21218, United States

[2] Research and Exploratory Development Department, Johns Hopkins University Applied Physics Laboratory, Laurel, Maryland 20723, United States

## Features definition

The parameters $\Delta S_{mix}$, $\Delta H_{mix}$, $\Omega$, $\eta$, $k_1^{cr}$, $\delta$, $\frac{E_2}{E_0}$, $\Delta\chi$ are defined in the Table S1.

**Table S1: Definition of the input features used in this study. The parameters are shown in the left column while their corresponding descriptions are shown in the right column.**

| Parameter | Description |
|---|---|
| $\Delta S_{mix} = -R \sum_{i=1}^{N} c_i \ln(c_i)$ | R is the gas constant which is equal to 8.314 J/(mol.T), and $c_i$ is the concentration of element i in atomic fraction. |
| $\Delta H_{mix} = \sum_{i=1,\ i\neq j}^{N} 4\Delta H_{i,j}^{mix} c_i c_j$ | $\Delta H_{i,j}^{mix}$ are calculated from available tables that are obtained from Miedema's model. |
| $\Omega = \frac{T_m \Delta S_{mix}}{|\Delta H_{mix}|}$ | $T_m$ is calculated from $\sum_{i=1}^{N} c_i T_{mi}$ wherein $T_{mi}$ is the melting temperature of element i. |
| $\eta = \frac{-T_{ann} \Delta S_{mix}}{|\Delta H_f|}$ | $T_{ann}$ is estimated as $0.8 T_m$, and $\Delta H_f$ is the most negative binary mixing enthalpy for forming IM ( i.e. $H_{i,j}^{IM}$ ) that are reported in [1]. |

---

* Corresponding authors:
  Email addresses: ashargh1@jhu.edu (A. K. Shargh), jelawady@jhu.edu (J. A. El-Awady)

| | |
|---|---|
| $k_1^{cr} = \dfrac{(1 - \dfrac{0.4T_m \Delta S_{mix}}{|\Delta H_{mix}|})}{\dfrac{\Delta H_{IM}}{\Delta H_{mix}}}$ | $\Delta H_{IM}$ is mixing enthalpy for forming IM. |
| $\delta = 100 \times \sqrt{\sum_{i=1}^{N} c_i [1 - \dfrac{r_i}{\sum_{j=1}^{N} c_j r_j}]^2}$ | $r_i$ is the atomic radius of element i. |
| $\dfrac{E_2}{E_0} = \sum_{j \geq i}^{N} \dfrac{c_i c_j |r_i + r_j - 2\bar{r}|^2}{(2\bar{r})^2}$ | $\bar{r}$ is calculated from $\sum_{i=1}^{N} c_i r_i$. |
| $\Delta \chi = \sqrt{\sum_{i=1}^{N} c_i [\chi_i - \sum_{j=1}^{N} c_j \chi_j]^2}$ | $\chi_i$ is Electronegativity of element i. |

## Features distribution

The feature distribution of Ti-free alloys with BCC+B2+Laves phases as well as correctly and falsely predicted Ti-alloys with BCC+B2+Laves phases are shown in Figure S1 – S7.

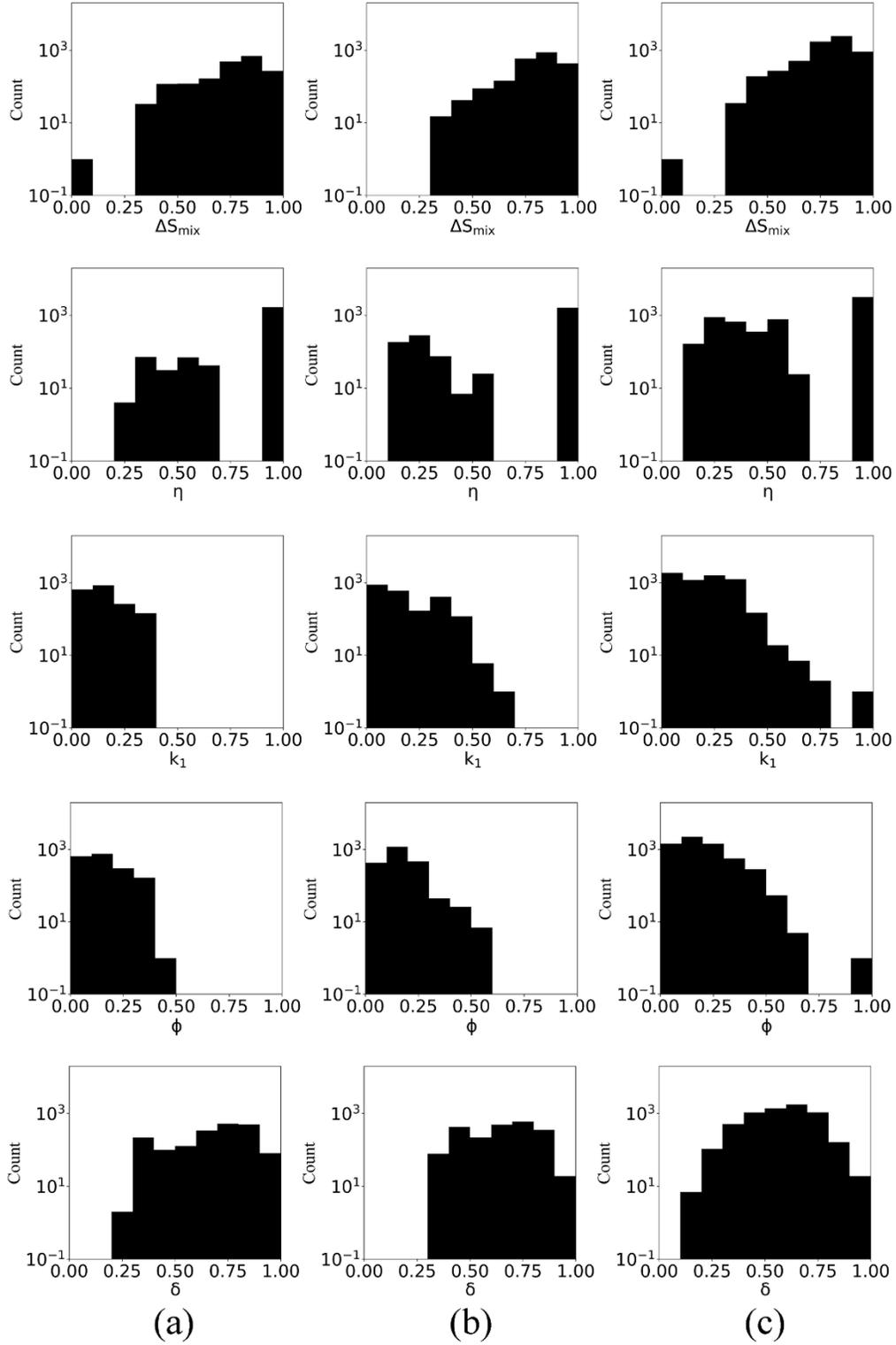

**Figure S1: Comparison of the feature distributions (range: 0-1) for alloys with BCC+B2+Laves phases. a) Normalized feature distributions for Ti-free. b) Normalized feature distributions for Ti-containing alloys correctly classified. c) Normalized feature distributions for Ti-containing alloys falsely classified. Five representative features of $\Delta S_{mix}$, $\eta$, $k_1^{cr}$, $\Phi$, and $\delta$ are shown.**

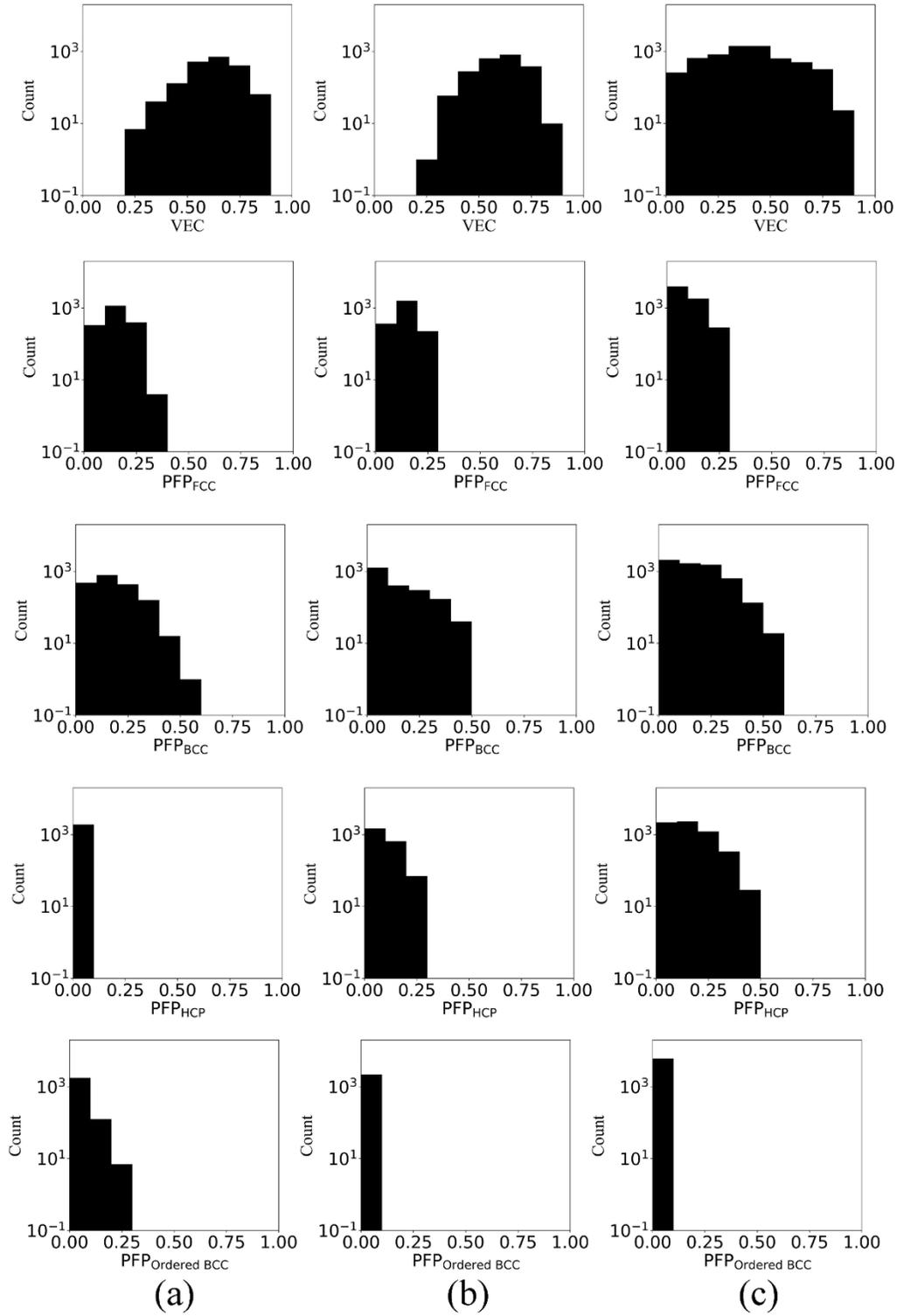

Figure S2: Comparison of the feature distributions (range: 0-1) for alloys with BCC+B2+Laves phases. a) Normalized feature distributions for Ti-free. b) Normalized feature distributions for Ti-containing alloys correctly classified. c) Normalized feature distributions for Ti-containing alloys falsely classified. Five representative features of VEC, $PFP_{FCC}$, $PFP_{BCC}$, $PFP_{HCP}$, and $PFP_{Ordered\ BCC}$ (i.e. $PFP_{B_2}$) are shown.

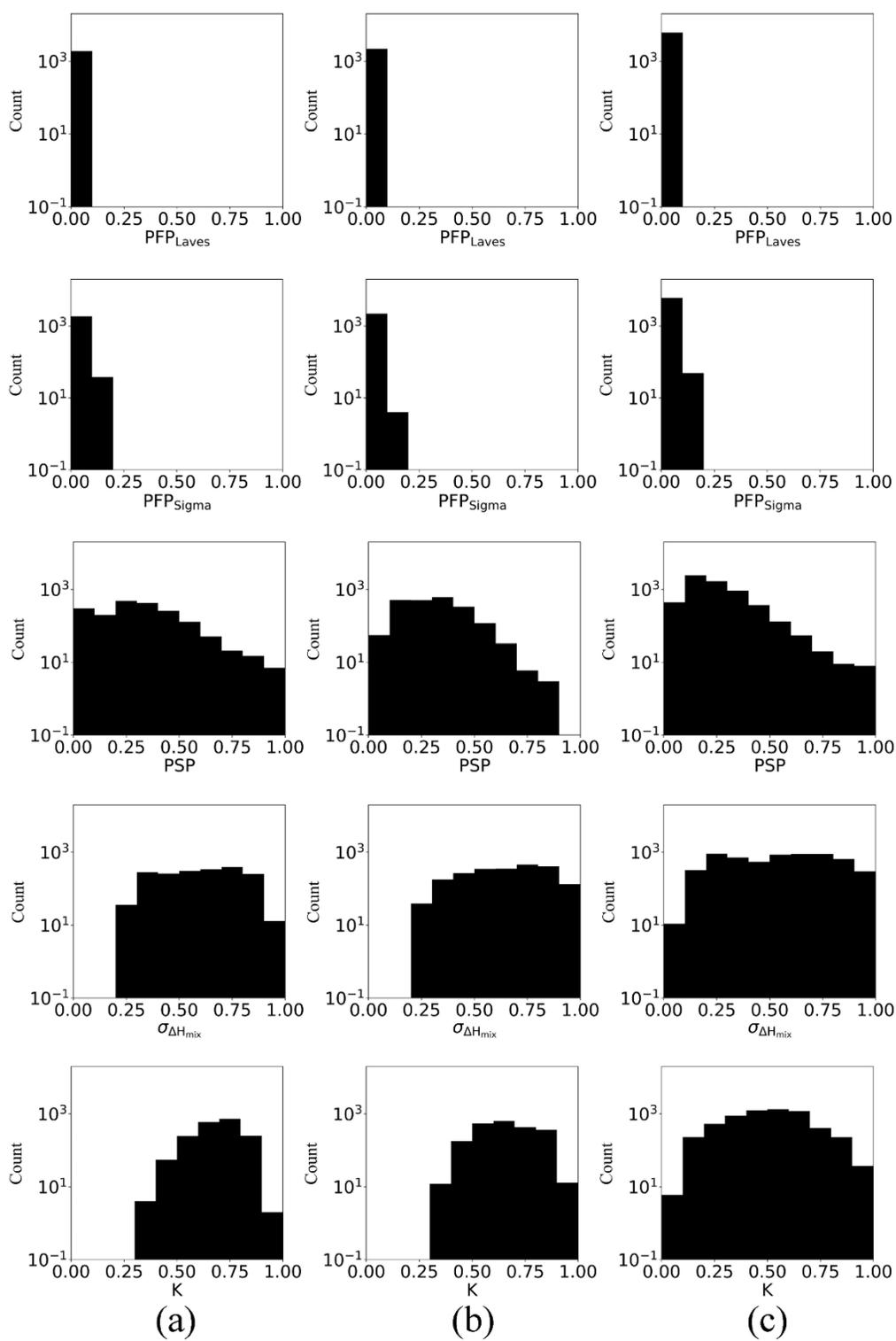

Figure S3: Comparison of the feature distributions (range: 0-1) for alloys with BCC+B2+Laves phases. a) Normalized feature distributions for Ti-free. b) Normalized feature distributions for Ti-containing alloys correctly classified. c) Normalized feature distributions for Ti-containing alloys falsely classified. Five representative features of $PFP_{Laves}$, $PFP_{Sigma}$, PSP, $\sigma_{\Delta H_{mix}}$, and K are shown.

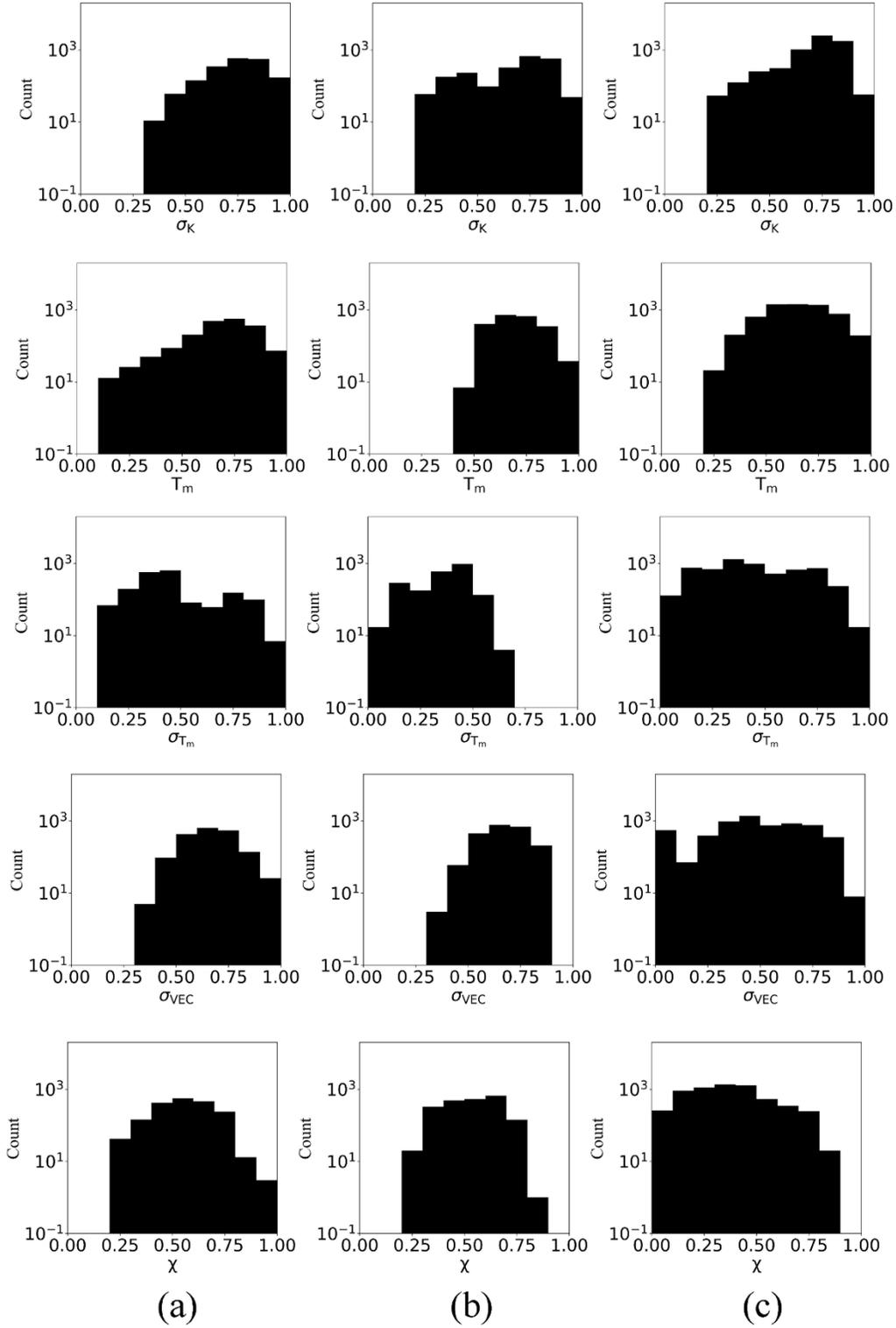

Figure S4: Comparison of the feature distributions (range: 0-1) for alloys with BCC+B2+Laves phases. a) Normalized feature distributions for Ti-free. b) Normalized feature distributions for Ti-containing alloys correctly classified. c) Normalized feature distributions for Ti-containing alloys falsely classified. Five representative features of $\sigma_K$, $T_m$, $\sigma_{T_m}$, $\sigma_{VEC}$, and $\chi$ are shown.

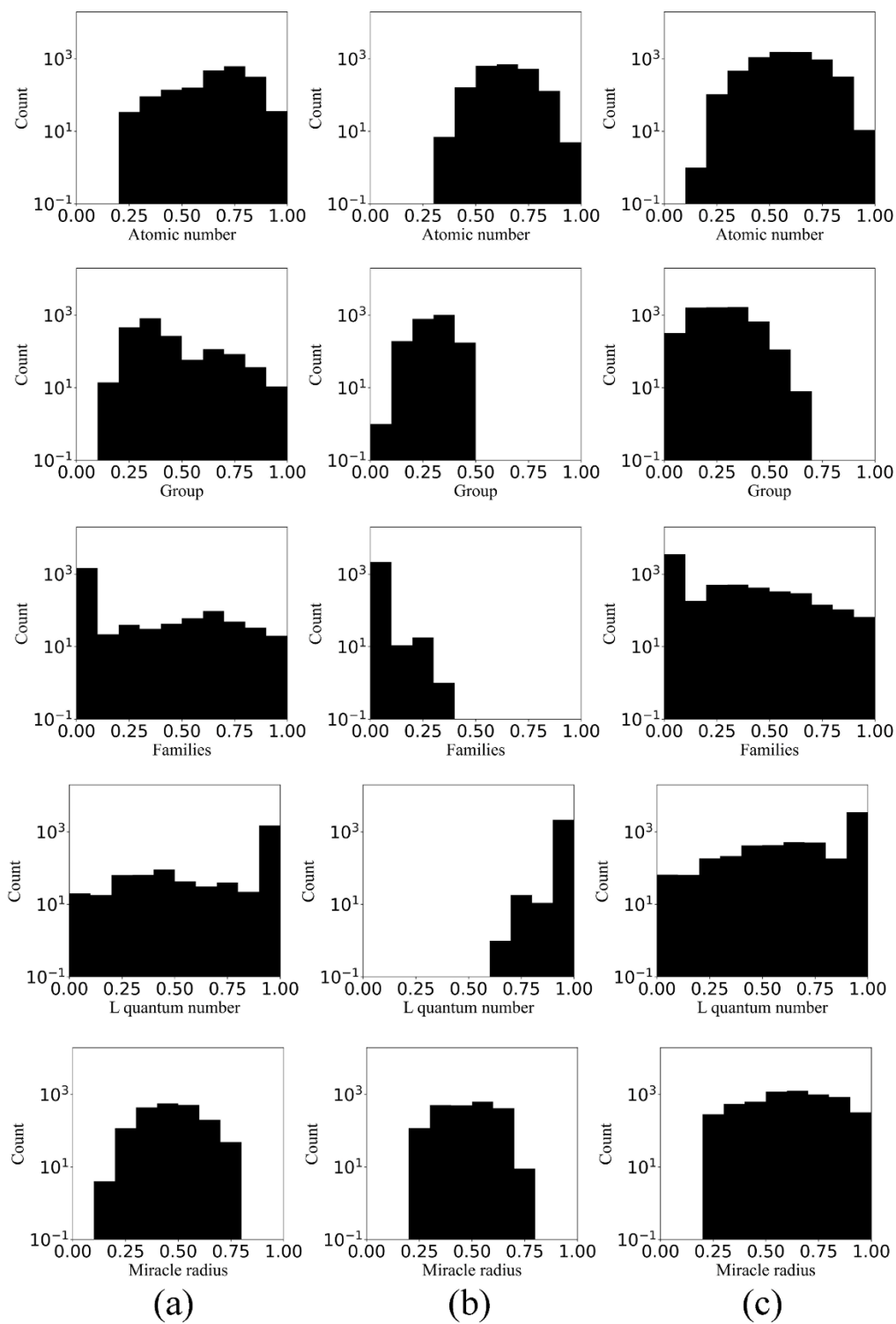

**Figure S5: Comparison of the feature distributions (range: 0-1) for alloys with BCC+B2+Laves phases. a) Normalized feature distributions for Ti-free. b) Normalized feature distributions for Ti-containing alloys correctly classified. c) Normalized feature distributions for Ti-containing alloys falsely classified. Five representative features of atomic number, group, families, L quantum number, and miracle radius are shown.**

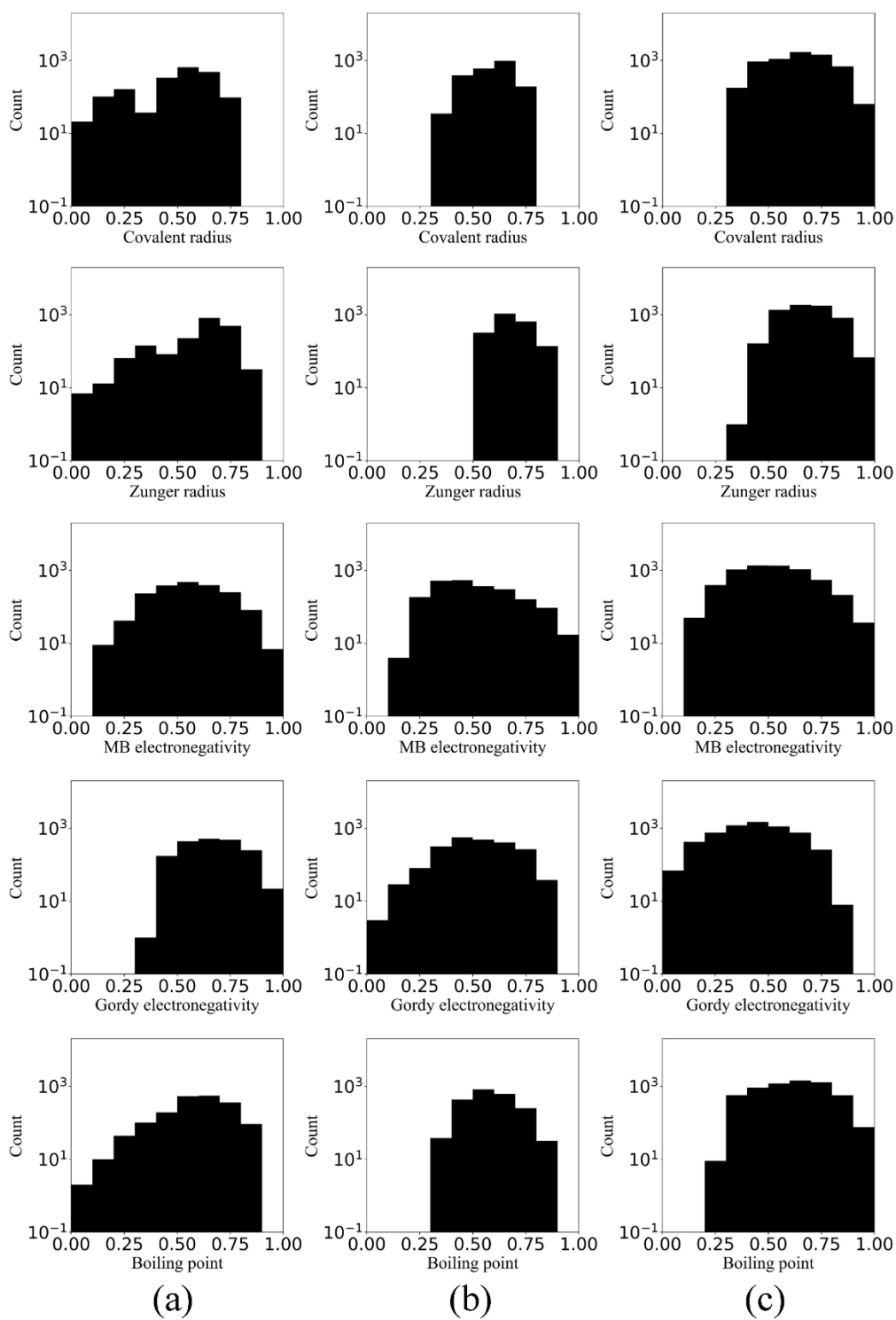

**Figure S6: Comparison of the feature distributions (range: 0-1) for alloys with BCC+B2+Laves phases. a) Normalized feature distributions for Ti-free. b) Normalized feature distributions for Ti-containing alloys correctly classified. c) Normalized feature distributions for Ti-containing alloys falsely classified. Five representative features of covalent radius, Zunger radius, MB electronegativity, Gordy electronegativity, and boiling point are shown.**

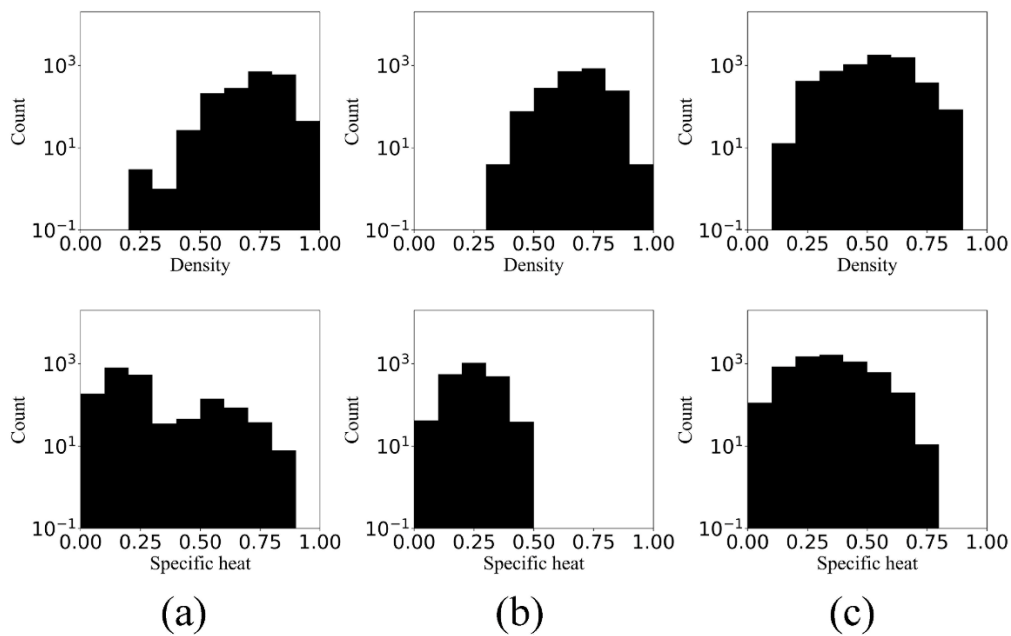

**Figure S7:** Comparison of the feature distributions (range: 0-1) for alloys with BCC+B2+Laves phases. a) Normalized feature distributions for Ti-free. b) Normalized feature distributions for Ti-containing alloys correctly classified. c) Normalized feature distributions for Ti-containing alloys falsely classified. Two representative features of Density, and Specific heat are shown.

## Supplementary references

[S1] M.C. Troparevsky, J.R. Morris, P.R. Kent, A.R. Lupini, G.M. Stocks, Criteria for predicting the formation of single-phase high-entropy alloys, Physical Review X 5 (2015) 011041.